\documentclass[aps,prl,reprint,groupedaddress]{revtex4-1}
\usepackage{graphicx}
\usepackage{amsmath}
\usepackage{amsfonts}
\usepackage{wasysym}
\usepackage{tabularx}
\usepackage{color}

\begin{document}

\title{Stability criterion for solitons of the ZK-type equations}
\author{E.A. Kuznetsov}
\affiliation{P.N. Lebedev Physical Institute, 53 Leninsky Ave.,
119991 Moscow, Russia\\
Novosibirsk State University, 2 Pirogov str., 630090 Novosibirsk, Russia;\\
L.D. Landau Institute for Theoretical Physics, 2 Kosygin
str., 119334 Moscow, Russia }

\begin{abstract}
Early results concerning the linear stability of the solitons in equation of
the KDV-type \cite{KUZNETSOV1984314} are generalized to solitons describing by  the ZK-type equation. The linear stability criterion for ground
solitons in the Vakhitov-Kolokolov form is derived for such  equations
with arbitrary nonlinearity. For the power nonlinearity the instability
criterion coincides with the condition of the Hamiltonian unboundedness from
below. The latter represents the main feature for appearance of collapse in
such systems.
\end{abstract}

\maketitle

\section{Introduction}

There are well known two the most popular multi-dimensional generalizations of the KDV equation:
the Kadomtsev-Petvishvili (KP) equation \cite{kadomtsev1970stability} and the so-called
Zakharov-Kuznetsov (ZK) equation \cite{zakharov1974threedimensional}. The KP equation belongs to the
universal models, it describes the nonlinear behavior of weakly nonlinear waves of the acoustic type.  In comparison with the KDV equation the KP equation takes
into account the diffraction of the  waves. In the
two-dimensional (2D) case, this equation admits integration by the inverse
scattering transform (IST) and therefore solitons in this case play very
essential role in the wave dynamics. For negative dispersion (KP-I) solitons
are one-dimensional KDV solitons propagating in any direction. They are
stable relative to the KP instability \cite{kadomtsev1970stability, zakharov1975instability}. For positive
dispersion (KP-II), in the 2D case, solitons are localized in both directions and have the
form of the so-called lump solutions \cite{manakov1977two}. These solitons realize
minimums of the Hamiltonian for the fixed momentum and by this reason are
stable in the Lyapunov sense \cite{kuznetsov1982two}. In the 3D case, however, 3D solitons
are unstable realizing saddle points of the Hamiltonian. In this case the
Hamiltonian turns out to be unbounded from below. As it was shown
numerically \cite{kuznetsov1983collapse, kuznetsov1986effect} due to the Hamiltonian unboundedness collapse
becomes possible in this case.

As far as the ZK equation concerns it describes ion-acoustic waves in
magnetized plasma with low $\beta $, which is the ratio between thermal
pressure and magnetic pressure. Among all MHD waves the ion-acoustic waves at $\beta\ll 1$
are the waves with the lowest frequencies and therefore they can not excite
another MHD waves while the nonlinear interaction between waves. Their group
velocity is mainly directed along the magnetic field. Dispersion of these
waves  is defined by the Debye radius along the magnetic field direction and
in the transverse direction by the ion Larmor radius. The ZK equation in the
dimensionless variables has the following form%
\begin{equation*}
u_{t}+\frac{\partial }{\partial x}\left[ \Delta u+3u^{2}\right] =0.
\end{equation*}%
This equation is written for waves propagating in one direction along the
magnetic field ($\parallel \widehat{x}$) with the sound velocity, the
Laplace operator is responsible for the wave dispersion, the nonlinear term
describes the nonlinear correction  to the sound velocity. This equation
belongs to the Hamiltonian type%
\begin{equation}
u_{t}=\frac{\partial }{\partial x}\frac{\delta H}{\delta u},\text{ \ }H=\int %
\left[ \frac{1}{2}\left( \nabla u\right) ^{2}-u^{3}\right] d\mathbf{r.}
\label{ZK-Hamiltonian}
\end{equation}%
Besides $H$ this equation conserves also the momentum $\mathbf{P}$ which $x$%
-component is positive definite quantity, $P=1/2\int u^{2}d\mathbf{r>}0$. The
simplest soliton solutions are stationary localized waves propagating with
velocity $V>0$ along the magnetic field, $u=u(x-Vt,\mathbf{r}_{\perp }).$
These solutions are stationary points of the Hamiltonian for fixed $P,$%
\begin{equation}
\delta (H+VP)=0.  \label{var}
\end{equation}%
As it was shown first time in \cite{zakharov1974threedimensional} (see also \cite{zakharov2012solitons}) ground\
(spherical symmetrical, without nodes) solitons realize minimum of the
Hamiltonian for fixed $P$ and therefore are stable in the Lyapunov sense.
This stability proof is based on the application of the Gagliardo-Nirenberg
inequalities following from the Sobolev embedding theorems. This result can
be easily generalized to the case of arbitrary power nonlinearity (see, e.g.
the review \cite{zakharov2012solitons})  when instead $u^{3}$ in the Hamiltonian stands $%
u^{p}$ with $p>2$.  The simple analysis, however, shows that the boundedness
of $H$ for fixed $P$ takes place at $p-2<4/d$ where $d$ is the space
dimension. For $p-2>4/d$ the Hamiltonian becomes unbounded from below that is
one of the criteria for the wave collapse (see \cite{kuznetsov1986soliton},\cite{zakharov2012solitons}) and therefore
we should expect instability of solitons in this case. When instead $u^{p}$
in the Hamiltonian $H$ we have arbitrary function $f(u)$ then the approach
developed in \cite{zakharov1974threedimensional} is not so effective. In this case, one needs to
consider the linearized problem.

The main aim of this paper is to generalize the results about linear soliton
stability in the equations of the KDV type \cite{KUZNETSOV1984314} to the equation
of the ZK type with arbitrary nonlinearity.  We show that the linear
stability analysis for solitons gives the Vakhitov-Kolokolov-type criterion
well known for the NLS equations \cite{vakhitov1973stationary}. For the power
nonlinearity the instability criterion coincides with the condition of the
Hamiltonian unboundedness from below.

\section{Linear stability problem}

Let us consider the equation of motion (\ref{ZK-Hamiltonian}) in the system
of coordinates moving with the velocity $V$ ($>0$) along $x$-axis which 
Hamiltonian is written as
\begin{equation*}
\widetilde{H}=\int \left[ \frac{1}{2}\left( \nabla u\right) ^{2}-f(u)\right]
d\mathbf{r+}VP.
\end{equation*}%
With respect to the function $f(u)$ we suppose that it vanishes for $%
u\rightarrow 0$ as $au^{2+\varepsilon }$ ($a,\varepsilon >0$) and increases
faster than $u^{2}$ as $u\rightarrow \infty $. Such a behavior guarantees
the existence of the soliton solutions $u=u_{s}(x-Vt,\mathbf{r}_{\perp })$
determined from the variational problem (\ref{var}):%
\begin{equation}
-Vu_{s}+\Delta u_{s}+f^{\prime }(u_{s})=0.  \label{stat}
\end{equation}%
We will consider only ground soliton solution of this equation which is
spherical symmetric and without nodes. The most important point is that this
solution is symmetric with respect to change $x\rightarrow -x$. Now let us
perform linearization of the equation of motion on the background on the
soliton solution putting $u=u_{s}(x,\mathbf{r}_{\perp })+w(x,\mathbf{r}%
_{\perp })$ where $w$ is a small perturbation. This results in the
following linear equation%
\begin{equation*}
w_{t}=\frac{\partial }{\partial x}\frac{\delta H^{\prime }}{\delta w}
\end{equation*}%
where $H^{\prime }$ is the second variation of the Hamiltonian $\widetilde{H}%
,$%
\begin{equation*}
H^{\prime }=\frac{1}{2}\int wLwd\mathbf{r\equiv }\frac{1}{2}\left\langle
w|L|w\right\rangle 
\end{equation*}%
and $L=-\Delta +V-f^{\prime \prime }(u_{s})$ is the Schroedinger operator.
Because $u_{s}$ is an even function relative to $x$ we decompose the
perturbation $w$ by even ($\varphi $) and odd ($\psi $) parts, $w=\varphi
+\psi $. As the result expansion of $H^{\prime }$ will contain two terms:%
\begin{equation*}
H^{\prime }=\frac{1}{2}\left( \left\langle \varphi |L|\varphi \right\rangle
+\left\langle \psi |L|\psi \right\rangle \right) ,
\end{equation*}%
Besides, these two functions $\varphi $ and $\psi $ are canonically
conjugated variables with the Hamiltonian $H^{\prime }:$%
\begin{equation*}
\varphi _{t}=\frac{\partial }{\partial x}\frac{\delta H^{\prime }}{\delta
\psi },\text{ \ }\psi _{t}=\frac{\partial }{\partial x}\frac{\delta
H^{\prime }}{\delta \varphi }.
\end{equation*}%
These linear equations should be implemented by the solvability condition%
\begin{equation}
\left\langle \varphi |u_{s}\right\rangle =0  \label{orthogonality}
\end{equation}
\ which is consequence of the conservation of $P.$

The soliton solution, evidently, is stable if both quadratic forms $%
\left\langle \varphi |L|\varphi \right\rangle $ and $\left\langle \psi
|L|\psi \right\rangle $ are of the same sign. It can be  easily seen that
the second quadratic part $\left\langle \psi |L|\psi \right\rangle $ is not
negative definite because $L\frac{\partial }{\partial x}u_{s}=0$, i.e.
function $\frac{\partial }{\partial x}u_{s}$ is a neutral eigen function of
the operator $L$ corresponding to a shift (along $x$) of the soliton as a
whole.   Because  $u_{s}$ is symmetric relative to $x$ and has no nodes, the
function $\frac{\partial }{\partial x}u_{s}$ corresponds to $p$-state.
According to the oscillatory theorem for the Schroedinger operators the $L$
operator will have  the ground eigen function with negative energy,
symmetric and without nodes. Among odd functions the function $\frac{%
\partial }{\partial x}u_{s}$  has the minimal energy $E=0$. Because any
(small) shift of the soliton as a whole can not influence on its stability
we can consider  the second quadratic form as positive definite.  

Now we turn to the question about a sign of the first quadratic part $%
\left\langle \varphi |L|\varphi \right\rangle $.  If it will be positive we
will get stability, and instability in the opposite case. Consider the eigen
value problem for the $L$ operator,%
\begin{equation}
L\varphi =E\varphi +Cu_{s}.  \label{eigen}
\end{equation}
Here we add the second term with constant $C$ which is  Lagrange multiplier
because of condition (\ref{orthogonality}).  Let us expand $\varphi $
through the eigen (even) functions of this operator ($L\varphi
_{n}=E_{n}\varphi _{n}$),  $\varphi =\sum_{n}C_{n}$ $\varphi _{n}.$ Hence
using the compatibility condition (\ref{orthogonality}) we arrive at the
following dispersion relation (compare with \cite{vakhitov1973stationary})%
\begin{equation*}
F(E)\equiv \sum_{n}\frac{\left\langle u_{s}|\varphi _{n}\right\rangle
\left\langle \varphi _{n}|u_{s}\right\rangle }{E_{n}-E}=0.
\end{equation*}%
Because of orthogonality $\left\langle u_{s}|\nabla _{\perp
}u_{s}\right\rangle =0$ in this sum there is absent a term with $E=0$.
Consider now behavior of the function $F(E)$ in the interval $E_{0}<E<E_{2}$
between  the energy of the ground state $E_{0}$ ($<0$) and the first
positive energy $E_{2}$ ($>0$).  If the  function will intersect the
abscissa axis at $E>0$ then  the quadratic form will be positive definite. In
this case evidently 
\begin{equation*}
\sum_{n}\frac{\left\langle u_{s}|\varphi _{n}\right\rangle \left\langle
\varphi _{n}|u_{s}\right\rangle }{E_{n}}<0.
\end{equation*}%
In the opposite case we have 
\begin{equation*}
\sum_{n}\frac{\left\langle u_{s}|\varphi _{n}\right\rangle \left\langle
\varphi _{n}|u_{s}\right\rangle }{E_{n}}>0
\end{equation*}%
when the quadratic form can take negative values. These sums can be
expressed in terms of the soliton solution if one differentiates equation
relative to $V$:%
\begin{equation*}
L\frac{\partial u_{s}}{\partial V}=-u_{s}.
\end{equation*}%
Hence on the class of functions orthogonal to  $u_{s}$ we can see that 
\begin{equation*}
\sum_{n}\frac{\left\langle u_{s}|\varphi _{n}\right\rangle \left\langle
\varphi _{n}|u_{s}\right\rangle }{E_{n}}=\left\langle
u_{s}|L^{-1}|u_{s}\right\rangle =-\frac{1}{2}\frac{\partial P}{\partial V}.
\end{equation*}%
Thus, if 
\begin{equation*}
\partial P/\partial V>0
\end{equation*}
soliton will be stable and unstable in the opposite case. This criterion
represents the analog of the Vakhitov-Kolokolov criterion\cite%
{vakhitov1973stationary}  for the NLS equations.  In the case of power
nonlinearity $f(u)=u^{p}$, the dependence of  momentum $P$  on $V$ turns out to be
powerful:  $P$ $\propto V^{\gamma }$ where 
\begin{equation*}
\gamma =\frac{2}{p-2}-\frac{d}{2}.
\end{equation*}
Hence one can see that the instability criterion for solitons $p-2>4/d$
coincides with the unboundedness condition of the Hamiltonian. Like for the
NLS-type equations we can state that the nonlinear stage of this instability
should result in the wave collapse.

\section{Conclusion}

Thus, we have found the linear stability criterion for ground soliton
solutions in the ZK-type equation.  This criterion is necessary and
sufficient:\ if $\partial P/\partial V>0$ the solitons are stable and
unstable in the opposite case. This criterion is analogous to the
Kolokolov-Vakhitov criterion for soliton stability in the NLS-type
equations. For power nonlinearity this criterion demonstrates different
behavior of the system. In the stable region solitons realize minimum of the
Hamiltonian with fixed momentum $P$, i.e. they are stable in the Lyapunov
sense. But it does not mean that scattering of solitons will be elastic.
While scattering of such solitons it is energetically favorable to form
solitons with higher amplitude. This process will be accompanied by radiation of small amplitude waves which play the role of friction in the system. For the systems with 
Hamiltonians unbounded from below  the nonlinear stage of the
soliton instability should result in the formation of singularity, probably,
in a finite time.

\section{Acknowledgments}

The  author thanks S. Roudenko who paid the author' attention on this unsolved
problem while the conference in Chernongolovka in the Landau Institute for
Theoretical Physics   at the end of this May. This work was supported by the
Russian Science Foundation (Grant No. 14-22-00174).


\begin{thebibliography}{99}

\bibitem{KUZNETSOV1984314}Kuznetsov  E.A. (1984),
{\it Soliton stability in equations of the KdV type},
Physics Letters A,
{\bf 101},
314 - 316 .

\bibitem{kadomtsev1970stability}
Kadomtsev B.B. and Petviashvili V.I. (1970),
{\it On the stability of solitary waves in weakly dispersing media},
Sov. Phys. Dokl,
 {\bf 15},
  {539--541}
  .

\bibitem{zakharov1974threedimensional} Zakharov V.E.  and Kuznetsov  E.A. (1974),
  {\it On threedimensional solitons},
  Zhurnal Eksp. Teoret. Fiz,
  {\bf 66}, 594--597.

\bibitem{zakharov1975instability} Zakharov V.E. (1975),
  {\it Instability and nonlinear oscillations of solitons},
  Soviet Journal of Experimental and Theoretical Physics Letters,
  {\bf 22},
  172.

\bibitem{manakov1977two} Manakov S.V.,  Zakharov V.E. ,Bordag L.A. , Its A.R., and Matveev  V.B. (1977),
  {\it Two-dimensional solitons of the Kadomtsev-Petviashvili equation and their interaction},
Physics Letters A,
  {\bf 63},
  205--206.
 
\bibitem{kuznetsov1982two} Kuznetsov E.A. and Turitsyn  S.K. (1982),
  {\it Two- and three-dimensional solitons in weakly dispersive media}, Sov. Phys. JETP,
  {\bf 55},
 844--847.



\bibitem{kuznetsov1983collapse} Kuznetsov E.A., Musher S.L., and Shafarenko A.V. (1983),
  {\it Collapse of acoustic waves in media with positive dispersion}, JETP Lett.,
  {\bf 37},
  241--245
  .

\bibitem{kuznetsov1986effect} Kuznetsov E.A.  and  Musher  S.L.(1986), {\it Effect of collapse of sound waves on the structure of collisionless shock waves in a magnetized plasma},
  Zhurnal Eksperimental’noi i Teoreticheskoi Fiziki,
  {\bf 91}, 1605--1619 .

\bibitem{zakharov2012solitons} Zakharov V.E. and Kuznetsov E.A. (2012), {\it Solitons and collapses: two evolution scenarios of nonlinear wave systems},
  Physics-Uspekhi,
  {\bf 55},
  535-- 556.
\bibitem{kuznetsov1986soliton}  Kuznetsov E.A.,Rubenchik A.M. and  V.E. Zakharov (1986), {\it Soliton stability in plasmas and hydrodynamics},
  Physics Reports,
  {\bf 142},
  103--165.
  
\bibitem{vakhitov1973stationary} Vakhitov N.G. and   Kolokolov A.A. (1973), {\it Stationary solutions of the wave equation in a medium with nonlinearity saturation}, Radiophysics and Quantum Electronics,
  {\bf 16},
  783--789.
 

\end{thebibliography}

\end{document}